\documentclass[aps,pre,onecolumn,epsf,graphicx,a4]{revtex4}
\usepackage[pdftex]{graphicx,color}    
\usepackage{amsfonts}
\usepackage{epsfig}
\usepackage{amsmath}
\usepackage{amssymb}

\def\pd2x{{\partial^2 \over \partial x^2}}

\usepackage{epsfig,latexsym}

\begin{document}

\title{\bf\noindent Random Walks between Leaves of  Random Networks}

\author{David Lancaster}

\affiliation{
Department of Computing and Mathematics, University of Plymouth,
Drake Circus, Plymouth PL4 8AA, UK.
}
\date{10 July 2012}
\begin{abstract}
We consider random walks that start and are absorbed on the
leaves of random networks and study the length of such walks.
For the networks we investigate, 
Erd\H{o}s-R\'enyi   random graphs and 
Barab\'asi-Albert scale free networks, 
these walks are not transient 
and we consider various approaches to
computing the
probability of a given length walk.
One approach is to label nodes according to both their total degree and the
number of links connected to leaf nodes,
and as a byproduct we compute  
the probability of a random node of a scale free network
having such a label.
\end{abstract}

\maketitle
\vspace{.2cm} 
\pagenumbering{arabic}

\section{Introduction}

Random walks have played a significant role in physics over the last century
and more recently their properties on random networks have received attention
as they can model transport
properties of complex systems. 
In this work we consider random walks that start on leaf nodes
of the random network and are absorbed whenever they reach another
leaf node. For certain types of network transport this is a natural restriction.
For example, while diffusion has been proposed to model communication
traffic on the internet \cite{internetdiffusion,internetdiffusion2,weightedinternet}, the majority of such traffic is between hosts
lying at the edge of the network and would be better modelled by
the type of random walk we consider in this paper. 

Past studies of non-absorbing random walks on random networks \cite{ReigerNoh,DenseER,Olivier,MFTER,MasudaKonno}
have often concentrated
on the mean first passage or  hitting times \cite{Redner}.
Typically, the mean first passage time scales with the
size of the network and the walk is transient in the thermodynamic limit.
In contrast, at least for the networks we have studied, we find that 
the probabilities of given path lengths between leaves are 
independent of the size of the network.
Moreover, the walks we study are often short and techniques
that rely on diffusion reaching equilibrium are not obviously adaptable
since the largest eigenvalue does not have time to become dominant.
Besides averaging over networks of a given class, we average over
many walks in order to find representative results for a given type of random network.
For a long running non-absorbing random walk, the equilibrium 
node occupation probability is proportional to the degree of that node
\cite{ReigerNoh}.
Even after averaging over walks,
this is not the the case for walks between leaves as links from leaf nodes must be treated
differently. 
One way to do this is to consider an 
approach based on labelling nodes 
according to both their total degree $k$ and the number of links
$l$ connecting to leaf nodes. 
We propose a technique in which the average occupation probabilities
are proportional to the number of non-leaf links 
besides adapting other methods that appear in the literature.

The networks we have chosen to study are the traditional
Erd\H{o}s and R\'enyi  (ER) \cite{RandomGraph} random graphs
and the scale free models due to Barab\'asi and Albert (BA)\cite{BarabasiAlbert}.
There is consensus that the scale free networks
model real networks better than ER random graphs.
On the other hand, random graphs are more analytically tractable,
principally due to their lack of correlation between node
degrees at each end of a link. Since it is essential that our networks have leaf nodes,
we work in the sparse regime of the ER graphs and choose the
simplest BA graph.

Our main interest lies in the probability, $p_t$, for a walk of length $t$ between leaves.
For short walks that can be enumerated,
this probability can sometimes be computed exactly 
and displays differences between even and odd length walks due to the possibility
of being absorbed by the originating leaf node. 
We expend some effort on these short walks, but are also interested in longer walks. 
For ER networks, the probability of longer walks decays exponentially but
the correlations and lack of homogeneity of BA networks destroys
this simple behaviour. 
Despite the hierarchical structure of the internet, data for the fraction of
packets arriving at an edge node after a certain number of hops 
displays an exponential decay over a range up to about 30 hops.
The most successful techniques to estimate the rate of decay 
of the probability for long walks are based on a modified assumption of
the equilibrium occupation of nodes labelled by 
$(k,l)$, the degree and the number of links to leaf nodes.


We commence with some simple and instructive network models
to develop intuition for the process. The next section covers 
Erd\H{o}s and R\'enyi  random graphs
and we provide both analytic and simulation results.
We then move on to discuss the problem on a 
scale free network: the $m=1$ BA model with leaves.
The conclusion returns to discuss the motivating example of the
internet and summarises our results.

\section{Simple Networks}

To commence, recall the random walk \cite{Feller}
on the half line in which the walker
starts by taking a step to the right from the origin, and is absorbed if it returns there. At each subsequent (discrete) time step
the probability of moving one step to the right is $p$ and that for a step to the left is $1-p$.
The recursion relation for the occupancy probability can be solved with the help of a Fourier
transform, and a generating function approach yields an expression for the first passage
probabilities \cite{Redner}. 
Paths that return to the origin always take an even number of steps and these paths
typically involve many backtracks or reversals. 
Counting these paths amounts to a
tree enumeration problem which is implicitly solved by  
the generating function computation.
The first passage probabilities, $r_t$,
can be expressed using a Catalan number as.
\begin{equation}
r_{2n}
= p^{t-1}(1-p)^n {(2n-2)!\over n! (n-1)!}
\end{equation}
The sum over probabilities for all length paths is related to the generating function
for Catalan numbers. This involves a square root and by taking the
appropriate sign we find that the overall probability of return is given by.
\begin{equation}
\sum_{n=1} r_{2n} = \left\{ 
\begin{array}{ll}
1& \mbox{$p  \leq 1/2$}\\
{1-p\over p}&\mbox{$p > 1/2$}
\end{array}
\right.
\end{equation}
So if the walk is biased to the right, the overall probability that it will ever return is
less than 1 and this behaviour is said to be transient. Of course, transience can only occur in
the limit of an infinitely large network.

Asymptotically, for long paths, the probability of a path of a given length
decays exponentially with the exponent $\gamma=-(1/2)\log 4p(1-p)$. 
\begin{equation}
\lim_{t\to\infty}r_t
= {1\over 4p\sqrt{\pi}}\left(4p(1-p)\right)^{t/2} = 
{1\over 4p\sqrt{\pi}}e^{-\gamma t} 
\end{equation}
The case $p=1/2$ is
special and gives rise to power law decay.

The case on a finite line with two boundaries or leaves is treated in \cite{GuptaSeth}, 
but this does not give rise to any new concepts.

\medskip

Now consider a Cayley tree. The degree of vertices is taken to be constant $k$ and rather
than follow the precise location of the random walk, we track the level
within the tree. We assume a Cayley tree with a large number of levels, but start from the 
boundary level which is the only level containing leaves.
The probability that the walk moves one level towards the root is $p=1/k$
while the probability it moves towards the boundary is $1-p$. 
In terms of levels, the rest of the analysis is identical to that of the 1D random walk just
considered, though in this case the real tree has many possible leaves
and walks that start and end on different nodes are accounted for.
For integer values of $k$, this walk will never be transient.
The exponent $\gamma = -(1/2)\log(4/k(1-1/k)$ does not depend on the size of
the graph, provided it is large enough.
Note that a non-absorbing random walk from the root of an infinite Cayley tree is transient with return 
probability $1/(k-1)$ for $k \ge 3$ \cite{Olivier}.

\medskip

The final simple network we consider is a modification of the random regular
network to allow the presence of leaves. This network is constructed by 
first taking an ordinary random regular network in which all nodes have the same degree
and links are connected along the lines of
the configuration approach of Molloy and Reed\cite{ReedMolloy}.
Then, to each node, an additional $l$ links are attached each ending in a leaf.
Each of the non-leaf nodes then has total degree $k$ with 
$l$ links to leaves and $k-l$ links to other non-leaf nodes.
From the regularity of the problem it is straightforward to see that the 
probability of a random walk starting and ending on a leaf and making
$t$ steps is.
\begin{equation}
p_t = {l\over k}\left({k-l\over k}\right)^{t-2} 
\end{equation}
Again the decay is exponential and the exponent $\gamma$ does
not depend on the size of the graph. By summing these probabilities we find that
the walk is never transient and that it has  mean length  $1+k/l$.

\medskip

This study of simple networks suggests that the probability of longer walks
decays exponentially 
for homogeneous networks and that the
exponent is not dependent on the size of the network.
Imagine that a walk has survived until step $t$, then the
probability, $\beta$, that it will survive one more step is just that of avoiding a leaf node.
\begin{equation}
\beta = e^{-\gamma} = 1 - P(\mbox{transition to a leaf node})
\label{roughexp}
\end{equation}
The decay is exponential when the transition probability does not depend on the length of the walk,
and this is the case for homogeneous networks. 
We will estimate the transition probability for ER and BA networks in the following
sections.


\section{Erd\H{o}s-R\'enyi  random graphs}

It is more awkward to analyse random walks on
networks in which the degree distribution is not fixed since the walk learns
information about the structure of the network instance as it proceeds and this memory
should be taken into account for backtracks. 
The most tractable models to consider are the
Erd\H{o}s and R\'enyi  \cite{RandomGraph}  (ER) random graphs 
due to both the absence of correlations between node degrees and  their
locally tree-like structure.
Indeed, since this is the natural model, random walks on ER graphs have been studied 
extensively; but only relatively recently has the hitting time,  which scales with the size of the graph,
been computed \cite{Olivier}.
One of the difficulties with traditional random walks on ER graphs has been that 
the interesting issues of transience only occur if the walk takes place on the giant component
which is only present above the percolation threshold at $\langle k\rangle=1$. 
Early work \cite{DenseER} avoided this difficulty by
working in the dense regime where with high probability, all nodes belong to the giant component.
A mean field like approach \cite{MFTER} allows accurate estimates of the mean first passage time,
but is more suited to numerical than analytic expression.

\begin{figure}[htbp]
\epsfig{file=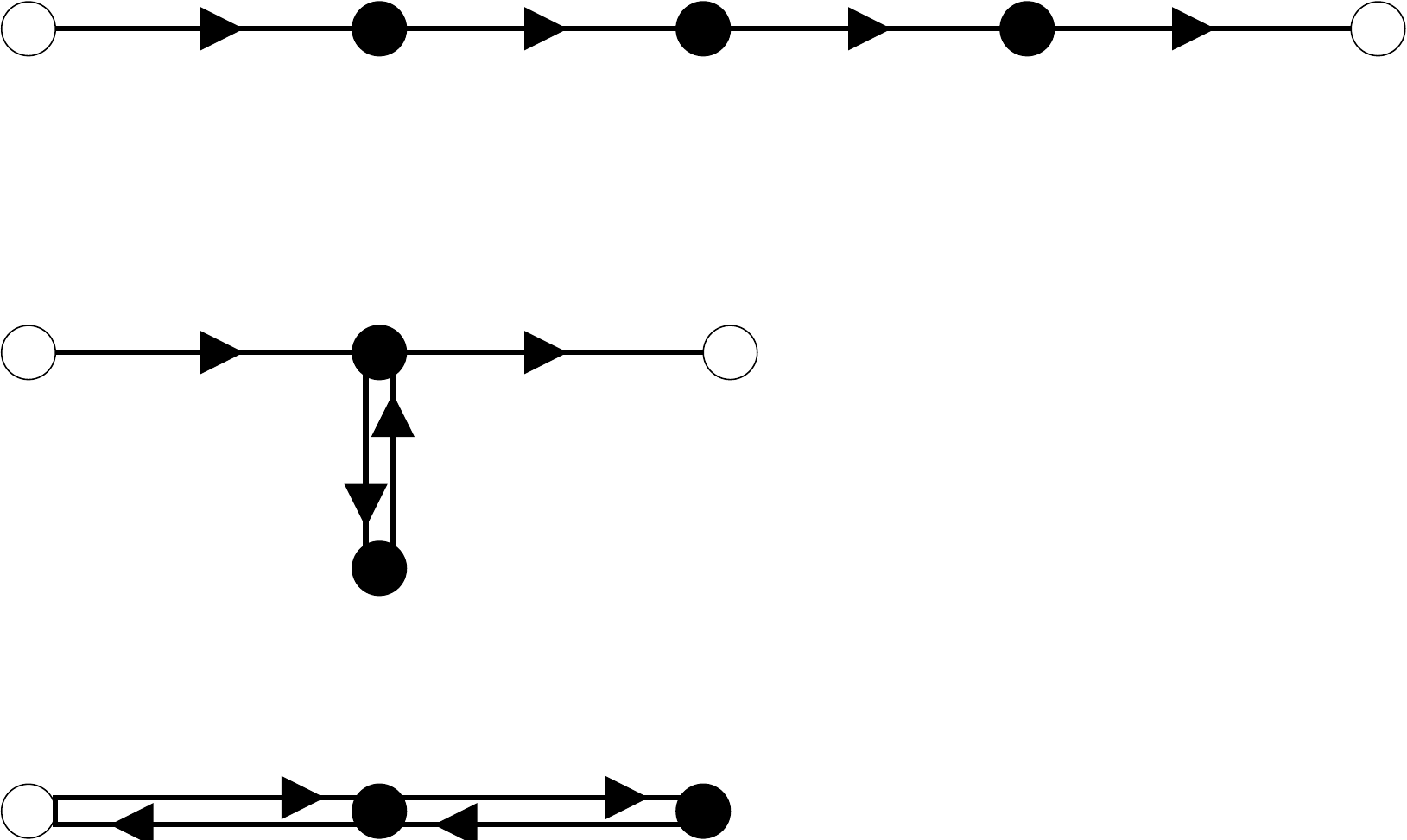,width=6cm,angle=0}
\caption{An enumeration of the three paths that contribute to a length 4 random walk between leaf nodes.
Open circles represent leaf nodes and filled circles are intermediate, non-leaf nodes. 
The contribution of each path is given in equation (\ref{eqfourstep}).}
\label{fourstep}
\end{figure}

In studying ER graphs we use the following notation.
The probability of a node picked at random having degree $k$ is Poisson distributed.
\begin{equation}
n_k = {\langle k\rangle^k\over k!} e^{-\langle k\rangle} 
\end{equation}
The probability that the node at the end of randomly chosen link has degree $k$ 
must include a factor for the number of links leading to that node.
\begin{equation}
 {k n_k \over \langle k\rangle} 
\end{equation}

Short paths can be enumerated, for example there are three $t=4$ length paths shown in
figure \ref{fourstep}.
By combining the probability of reaching a node with given degree with the random walk
factor for leaving it, these paths are seen to contribute respectively.
\begin{eqnarray}
\label{eqfourstep}
& &\left(\sum_{k=2} {k n_k\over \langle k\rangle} {(k-1)\over k}\right)^3 {n_1\over \langle k\rangle} \nonumber\\
& & \left(\sum_{k=2} { k n_k\over \langle k\rangle} {1\over k}\right)
\left(\sum_{k=2} {k n_k\over  \langle k\rangle} {(k-1)(k-2)\over k}\right){n_1\over \langle k\rangle}\\
& &\left(\sum_{k=2} {k n_k\over \langle k\rangle} {1\over k}\right)\left(\sum_{k=2} {k n_k\over  \langle k\rangle} {(k-1)\over k^2}\right)\nonumber
\end{eqnarray}
The sums can be written in terms of exponential integrals and are used to check
numerical simulations. 
This  diagrammatic approach  is restricted to very short paths as
no simple recursion exists and the number of diagrams increases rapidly.

\begin{figure}[htbp]
\epsfig{file=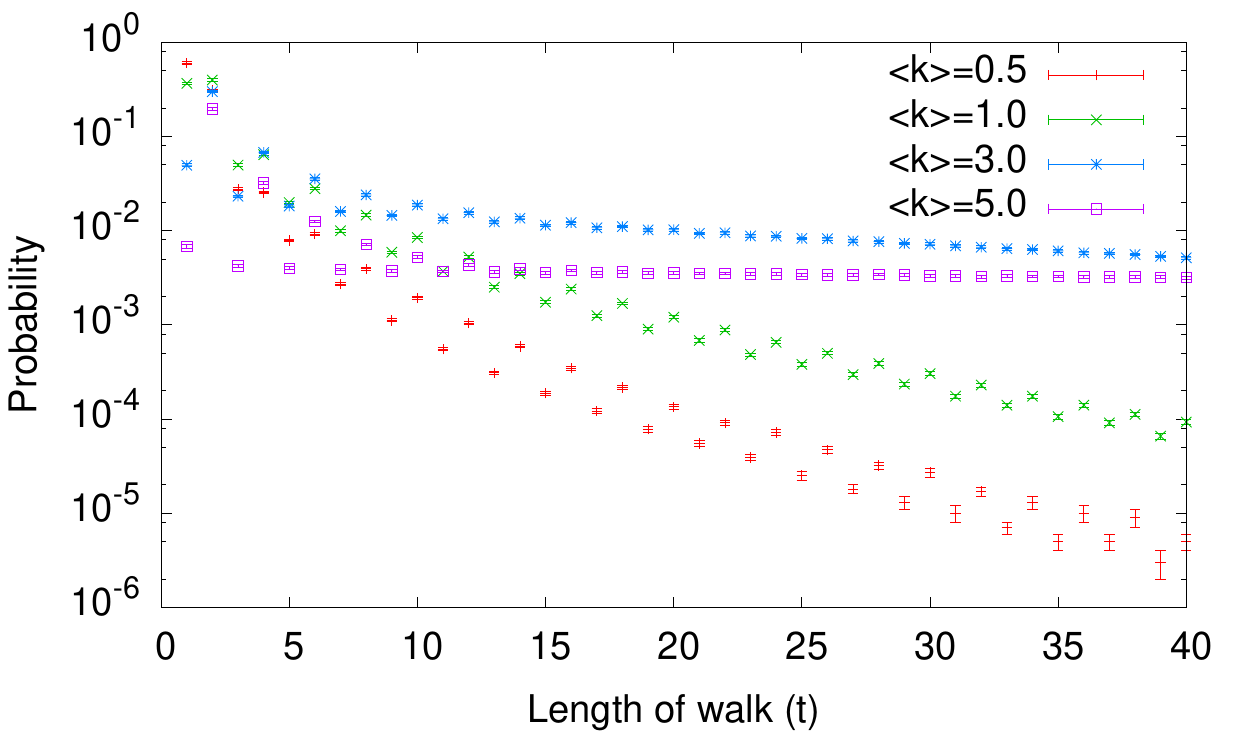,width=10cm,angle=0}
\caption{The probability of a random walk between leaves of length $t$ on an ER graph. Shown for 
various values of $\langle k \rangle$, the mean degree .}
\label{erdecay}
\end{figure}

\begin{figure}[htbp]
\epsfig{file=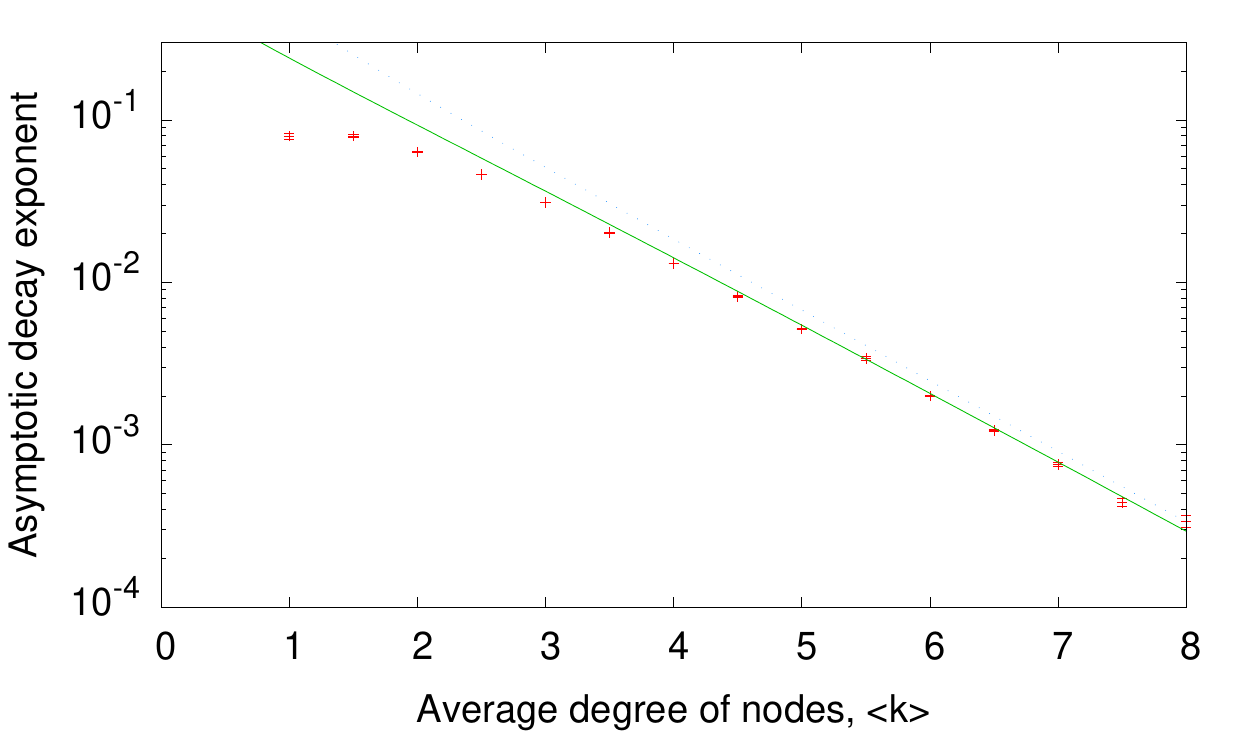,width=10cm,angle=0}
\caption{The decay exponent for random  walks between leaves on an ER graph
shown for various values of the mean degree parameter. The dotted line
shows the rough estimate of the the exponent from equation (\ref{ERroughexp}).
The continuous line is the result of a more sophisticated argument given at the end of  section \ref{ERkl}.}
\label{erscale}
\end{figure}

Numerical studies indicate that the walks between leaves that we consider in this paper 
have the same properties as on the simple models: they
do not scale with the size of the system and their probability decays as 
shown in figure \ref{erdecay}.
For these leaf random walks there is
no need to restrict the starting leaf nodes to be part of the giant component
and we include the contribution from walks on the finite clusters.
Below the percolation transition, where only finite clusters exist, the decay suffers strong  finite
size effects and is not exponential. 
Above the transition, the decay is exponential and
in figure \ref{erscale}
we show the exponent as a function of the mean degree parameter $\langle k\rangle$ of the model.
These numerical results are obtained from following 10000 random walks on each of 100 graphs of size 8000.
The number of random walks followed must be large both at small and large $\langle k\rangle$. At small
$\langle k\rangle$ there are many possible leaf nodes to start from, though the walks tend to be short.
At large $\langle k\rangle$ there are not so many leaf nodes, but the walks can take many different paths.
These results do not change as the network size is increased suggesting that transience does
not occur.

The behaviour of the exponent at large $\langle k\rangle$  can be estimated 
following the argument in equation (\ref{roughexp}).
For large $\langle k\rangle$  we ignore memory effects and estimate 
the probability that a walk will survive one more  step  as.
\begin{equation}
\beta = \sum_{k=2} {k n_k \over \langle k\rangle} = 1- {n_1\over \langle k\rangle} =1 - e^{-\langle k\rangle}
\label{ERroughexp}
\end{equation}
The exponent is thus $\gamma =-\log(1-e^{-\langle k\rangle})$ 
or approximately $e^{-\langle k\rangle}$ for large $\langle k\rangle$. This estimate is shown in
as a dotted line in figure \ref{erscale} and indeed matches the simulations for large  $\langle k\rangle$.
In a later section we improve this estimate.

We now proceed to discuss more sophisticated approximations, with the
aim of improving the prediction of walk survival probabilities in the region
where the exact enumeration in not feasible, but the exponential behaviour has
either not taken over, or the  estimate above is not accurate.

\subsection{Labeling according to ($k,l$)\label{ERkl}}

Since the identification of leaf nodes is crucial for the random walks we study in this
paper, we first propose an approach based on labelling each node according to both its overall
degree $k$ and the number of links to leaf nodes $l$. For ER graphs it is straightforward to 
write the probability, $q_{kl}$, that a node chosen at random has label $(k,l)$.
\begin{equation}
q_{kl} = {k\choose l} e^{-l \langle k\rangle}\left( 1-e^{-\langle k\rangle}\right)^{k-l} n_k
\end{equation}
The sum over leaf links of course obeys.
\begin{equation}
\sum_{l=0}^k q_{kl} = n_k
\end{equation}
The expectation value for the number of links to leaves found on a randomly chosen node
is:
\begin{equation}
\langle l\rangle  = \sum_k\sum_l lq_{kl} = n_1 = \langle k\rangle e^{-\langle k\rangle}
\end{equation}
In order to proceed, we first identify the probabilities with which a walk will find a $(k,l)$ node.
The probability, $v_{kl}$, that the link from a randomly chosen leaf node connects to a $(k,l)$ node is:
\begin{equation}
v_{kl}={l q_{kl} \over \langle l\rangle } = {l q_{kl} \over \langle k\rangle e^{-\langle k\rangle}} 
\end{equation}
Similarly, the probability, $w_{kl}$, that a randomly chosen link from a non-leaf node connects to a $(k,l)$ node is:
\begin{equation}
w_{kl}={(k-l) q_{kl} \over \langle k\rangle-\langle l\rangle}= {(k-l) q_{kl} \over \langle k\rangle(1-e^{-\langle k\rangle})} 
\label{eqnwkl}
\end{equation}
It is now possible to write down probabilities for given length paths using the simplest form of approximation in which
there is no memory and backtracks are not accounted for. 
The separate contributions at each step of the walk from the transition probability to a given degree $(k,l)$ node 
and from the random walk are clearly exposed.
%
%
%
\begin{eqnarray}
p_1 &=& v_{11} = e^{-\langle k\rangle}\nonumber\\
p_2 &=& \sum_{k=2}\sum_{l=1}^k v_{kl}{l\over k} = 
{\left(1-e^{-\langle k\rangle}\right)^2\over \langle k\rangle}\nonumber\\
p_t &=& \sum_{k=2}\sum_{l=1}^k v_{kl}{(k-l)\over k} 
\left[ \sum_{k=2}\sum_{l=1}^k (k-l) w_{kl}{(k-l)\over k} \right]^{t-3}
\sum_{k=2}\sum_{l=1}^k (k-l) w_{kl}{l\over k}\\
& =& 
{e^{-\langle k\rangle}(1-e^{-\langle k\rangle})(1-\langle k\rangle-e^{-\langle k\rangle})^2\over \langle k\rangle^2}
\left[1- {e^{-\langle k\rangle}(2\langle k\rangle-1+e^{-\langle k\rangle})\over \langle k\rangle}\right]^{t-3}\nonumber
\end{eqnarray}
Where the final line is valid for any $t \geq 3$.
The predictions for paths of length 1 and 2 match the exact enumeration results 
and that for length 3 differs from the exact result by  a factor of $(1- e^{-\langle k\rangle})$.
However, the prediction contained in the last equation above, 
for the exponent of the decay is less successful. In the limit of large $\langle k\rangle$ 
this prediction differs by a factor a 2 from equation (\ref{ERroughexp})
which suggested a value of $e^{-\langle k\rangle}$
and according to figure  \ref{erscale} the earlier argument was correct. 
Moreover the failure extends to transience as the sum of probabilities is
1/2 in the large $\langle k\rangle$ limit which does not match numerical expectations.
Given these inadequacies, we do not show a figure with the predictions from this approximation
and conclude that while neglecting both memory effects and backtracks is possible
for very short paths, it fails for longer paths even at large $\langle k\rangle$.

We can reproduce the simple prediction given in equation (\ref{ERroughexp})
for the asymptotic decay exponent using the $(k,l)$ 
formalism. Assuming that after some steps, the walk is on a node of unspecified degree, then
according to equation (\ref{eqnwkl})
the probability that its next step takes it to a node with label $(k,l)$ is $w_{kl}$.
In this case the 
probability that the walk is not absorbed in the next step is.
\begin{equation}
\beta = \sum_{k=2}\sum_{l=0}^k w_{kl} = 1 - e^{-\langle k\rangle}
\end{equation}
So this argument yields precisely the same exponent as the rough one.
We can obtain a better estimate of the exponent by assuming that when averaged over many leaf walks, the
occupation probabilities  take values similar 
to those attained by an non-absorbing walk in equilibrium. For traditional walks the equilibrium 
occupation probabilities are proportional to the number of incoming links: $k n_k/\langle k\rangle$ \cite{ReigerNoh}.
For leaf walks the natural generalisation of this equilibrium occupation probability is proportional to 
the number of non-leaf connections.
So the $w_{kl}$ not only represent 
transition probabilities, but when properly normalised can represent the occupation
probabilities of $(k,l)$ nodes  for $k\geq 2$.
Numerical investigations that track which sites are visited suggest that such an assumption is valid in the large $\langle k\rangle$ limit.
With the appropriate normalisation to ensure the correct sum of occupation probabilities over non-leaf nodes,
the estimate of the occupation probability of a  $(k,l)$ node is.
\begin{equation}
{w_{kl}\over (1-e^{-\langle k\rangle})}
={(k-l) q_{kl} \over \langle k\rangle(1-e^{-\langle k\rangle})^2} 
\end{equation}
The probability of stepping from a $(k,l)$ node to another non-leaf node is simply $(k-l)/k$, so overall the
probability of surviving one step is.
\begin{equation}
\beta = {1 \over \langle k\rangle(1-e^{-\langle k\rangle})^2} \sum_{k=2}\sum_{l=0}^k 
{(k-l)^2 \over  k}q_{kl}
=
1 - {e^{-\langle k\rangle}\left( {\langle k\rangle}-1 + e^{-\langle k\rangle}\right)
\over \langle k\rangle(1-e^{-\langle k\rangle})}
\label{ERklbetterexp}
\end{equation}
In the limit of large $\langle k\rangle$  this reproduces the rough result for the exponent
in equation (\ref{ERroughexp}),
but as can be seen from the continuous line in figure  \ref{erdecay}, this
prediction matches the simulation results slightly better.

\subsection{Memoryless Approximation\label{ERKN}}

Since the paths are typically short compared to the size of the network, 
the spirit of this paper is  to use methods that approximate the path enumeration
to make it tractable.
An interesting, though flawed, paper in this context is that of Masuda and Konno \cite{MasudaKonno}.
These authors write recursion formulae for the return probabilities
of a non-absorbing walk  to an arbitrary node
using an approximation that takes memory into account within each backtrack,
but regards each separate backtrack as a new path.
The approximation is expected to be successful at large $\langle k \rangle$.
Unfortunately the paper \cite{MasudaKonno} does not take into account the factor of the number of links that weights
the probability of finding a certain degree node at the end of a randomly chosen link, and this
compromises their results. Nonetheless the defect is simply remedied and
their approach can be modified for our random walks between leaves. 

\begin{figure}[htbp]
\epsfig{file=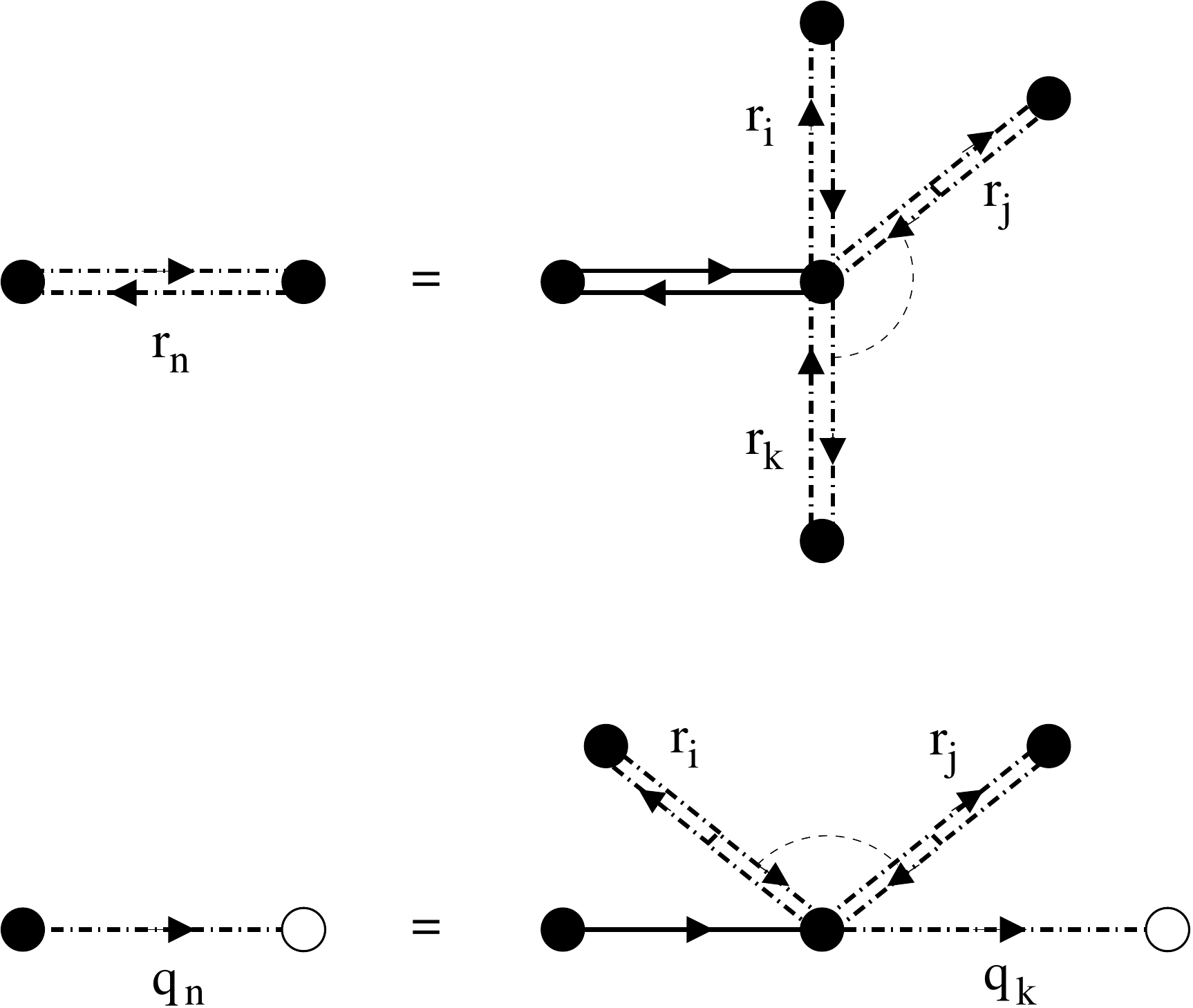,width=8cm,angle=0}
\caption{Diagrammatic representation of  the recursion used in the memoryless approximation.
The top diagram is for the first passage probability without absorption, $r_n$, and the lower
diagram is for the probability of being absorbed at some other leaf, $q_n$. The dashed
lines represent these quantities and the solid lines are single steps between nodes.
The open circles are leaf nodes and the solid circles are non-leaf nodes.
The sum over partitions is represented with the dotted arc.}
\label{KNrecursion}
\end{figure}

By considering the top diagram shown in figure \ref{KNrecursion}
the probability of first passage to a node after $2t$ steps without being absorbed, $r_t$, 
obeys the following recursion.
\begin{equation}
r_{t+1} = \sum_{a=1}^t m_{a+1}
\sum_{\{t_i\} \sum t_i=t}
\prod_{i=1}^a r_{t_i}
\label{rcoeffrecurse}
\end{equation}
The sum is over all integer partitions of $t$ including all possible orderings of those partitions.
The significance of the
partitions is that they represent the length of each backtrack shown in figure \ref{KNrecursion}.
In equation (\ref{rcoeffrecurse}), this sum over partitions 
is decomposed into a sum over the number of lobes or backtracks in the diagram 
(corresponding to the order of the partition),
and a sum over partitions into precisely this many lobes.
The parameters $m_a$ are correctly given by.
\begin{equation}
m_a = \sum_{k=2} {n_k \over \langle k\rangle} \left(1-{1\over k}\right)^{a-1}
\end{equation}
where the lower limit of the sum imposes the condition of no absorption and this is the
constitutes the only difference in computing $r_t$ with \cite{MasudaKonno}.
The form (\ref{rcoeffrecurse}) illustrates the nature of the approximation as the memoryless
nature consists in regarding each walk on a lobe as independent.

A generating function approach is useful and we write for
these first passage probabilities without absorption.
\begin{equation}
R(x) = \sum_{t=1} r_t x^t
\end{equation}
Later, when paths that are not of even length are taken into account, we will need to consider $R(x^2)$. 
In terms of $R(x)$ the recursion takes on a more convenient form than equation (\ref{rcoeffrecurse}).
\begin{equation}
R(x) = x \sum_{a=0} m_{a+1} \left(R(x)\right)^a
\label{Rrecurse}
\end{equation}

The probability that a walk lasts $t$ steps before being absorbed by a different
leaf node, $q_t$, also obeys a recursion relation shown diagrammatically in the lower part of
figure \ref{KNrecursion}. The recursion is slightly different for even and odd values of
$t$, but can be combined and is conveniently expressed in terms of the generating function.
\begin{equation}
Q(x) = \sum_{t=1} q_t x^t
\end{equation}
%
%
\begin{equation}
Q(x) = q_1 x + x Q(x) \sum_{a=0} s_{a+2} \left(R(x^2)\right)^a
\end{equation}
The first coefficient is $q_1 = n_1/\langle k\rangle$ which for ER graphs is
simply $e^{-\langle k\rangle}$.
The parameters $s_a$ are defined as.
\begin{equation}
s_a = \sum_{k=2} {k n_k \over \langle k\rangle} \left(1-{1\over k}\right)^{a-1}
\end{equation}
In some cases, it could be argued that the lower limit of this sum ought to be increased, but 
to keep track of these cases would be against the spirit of the approximation.
For ER graphs, the parameters $s_a$ and $m_a$ are related by.
\begin{equation}
s_a = s_{a-1} - m_{a-1}
\label{smreln}
\end{equation}

While Masuda and Konno invert equation (\ref{Rrecurse}) to obtain explicit
expressions for the first passage probabilities in terms of sums over partitions \cite{MasudaKonnoNote},
we have found using the recursion relations themselves to be numerically convenient and
speedy. Using algebraic manipulation software, the equation for the
generating function (\ref{Rrecurse}) conveniently encodes all the separate
equations for each coefficient and can be solved quickly and accurately.
The solution for $R(x)$ can then be used to provide values for the 
probabilities $q_t$ using the equation below.
The combined probability of absorption after $t$ steps is the sum of the 
contributions from $r_t$ and $q_t$
where the first term only contributes for even length paths. 
\begin{eqnarray}
Q(x) &= &{q_1 x\over 1- x \sum_{a=0} s_{a+2}  \left(R(x^2)\right)^a}
= q_1 x  \sum_{j=0}\left[x \sum_{a=0} s_{a+2}  \left(R(x^2)\right)^a\right]^j\\
&= &{q_1 x \left(1-R(x^2)\right)\over 1- x \left(s_1 -R(x^2)/x^2\right)}
= q_1 x  \sum_{j=0}\left[{x \left(s_1 -R(x^2)/x^2\right)\over \left(1-R(x^2)\right)}\right]^j
\end{eqnarray}
The second line is specific to ER graphs as it uses the relationship between parameters in equation (\ref{smreln}).
At $x=1$ this equation simplifies and by 
using the explicit forms of $q_1 = e^{-\langle k\rangle}$ and $s_1=1 - e^{-\langle k\rangle}$ we find 
$R(1)+Q(1)=1$. So at the level of this approximation, leaf walks are never
transient on ER graphs. Since the approximation is expected to improve as
$\langle k\rangle$ increases, and this is the regime in which transience
is most likely to occur, the result provides strong evidence that transience does not occur.
The result for the full generating function is 
helpful as it avoids the need to calculate the parameters $s_a$, which is
otherwise the most time consuming part of the solution.
The alternative approach of solving the generating function $R(x)$ by
iteration and then obtaining the coefficients from residues, works for
small $t$, but becomes slow and unstable at larger $t$. 

\begin{figure}[htbp]
\epsfig{file=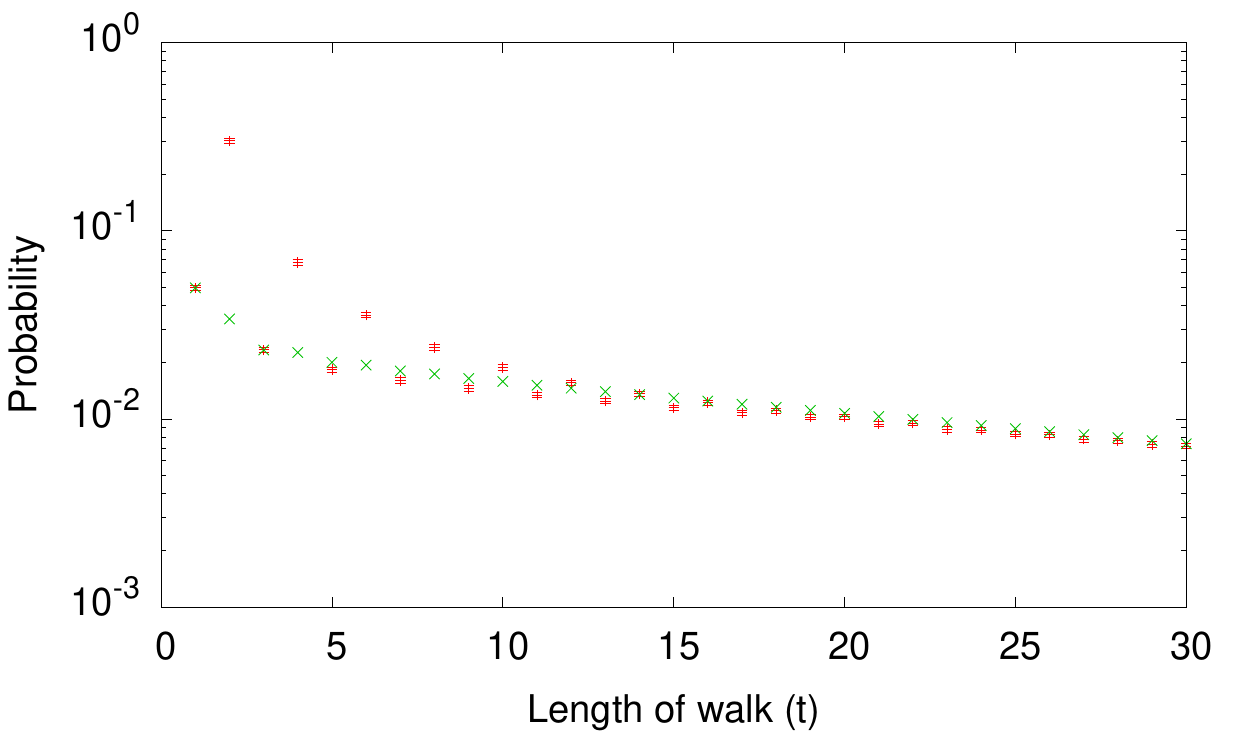,width=10cm,angle=0}
\caption{The probability for random  walks between leaves on an ER graph
shown for $\langle k\rangle = 3$.  Points with error bars are from
simulation and the crosses are predictions from the memoryless approximation.}
\label{knerz3}
\end{figure}

Figure \ref{knerz3} shows the predictions of this memoryless approach against the
results of simulations for a value of $\langle k\rangle = 3$. Although this value is
not very large, agreement is reasonable.
The first three points are identical to the
exact prediction but later points tend to track the curve for odd length walks.
At larger values of $\langle k\rangle$ the agreement is even better.
At $\langle k\rangle = 1$, the predictions are successful for $t<10$,
but for longer paths, the theory fails to capture the observed behaviour.

This approximation provides an improvement over the techniques used in the previous section as it
contains information about backtracking paths. For smaller values of $\langle k\rangle$,
where backtracks and memory effects become more important the approximation fails.



\section{Scale Free Networks}

Scale free networks (BA) \cite{BarabasiAlbert} created though a growth process involving
preferential attachment provide a popular alternative model of random networks.
We consider the simplest version of this system which according
to the classification of  \cite{BarabasiAlbert} is the $m=1$ model with preferential attachment 
and with an initial network
consisting of a pair of nodes connected by a link. The growth process leads to a
directed network with $\langle k\rangle=2$ and with $2/3$ of the nodes being leaves.
The degree distribution is given by \cite{KrapivskyRedner}.
\begin{equation}
n_k = {4\over k(k+1)(k+2)}
\end{equation}
Though if we take into account the age of nodes, older nodes tend to have larger degree 
while most leaf nodes are young.
In general, these networks suffer from strong finite size effects \cite{BAFiniteSize}.

Due to the way in which they are created, this kind of scale free network
is not homogeneous, has
strong correlations and in these respects is very different from the ER networks.
The directed nature of the network is crucial in understanding the correlations
The fraction of nodes of degree $k$ that attach to an ancestor node of degree
$l$ (the degree of the ancestor, $l$, must be greater or equal to 2), is \cite{KrapivskyRedner}
\begin{equation}
e_{kl} = {4(l-1)\over k(k+1)(k+l)(k+l+1)(k+l+2)}
+{12(l-1)\over k(k+l-1)(k+l)(k+l+1)(k+l+2)}
\label{BAcorrelate}
\end{equation}
%
%
%
Random walks on this network have no knowledge of the direction of the links, but
a full analysis of the walk should take this into account.

\begin{figure}[htbp]
\epsfig{file=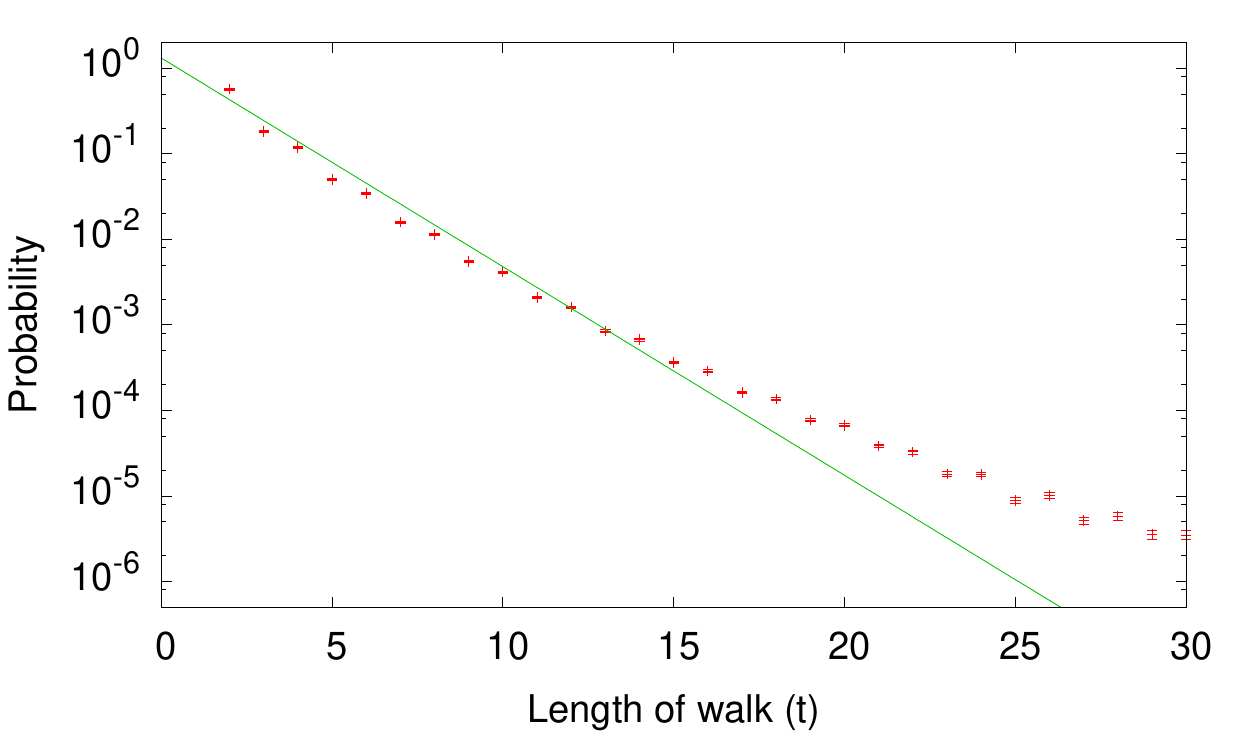,width=10cm,angle=0}
\caption{The probability for random  walks between leaves on a BA graph.
Points with error bars are from
simulation based on 20000 walks on each of 100 graphs of size 100000.
The continuous line is a prediction based on labelling nodes and is explained in the text.}
\label{bawalk}
\end{figure}

Numerical results are shown in figure \ref{bawalk}. There is no evidence for transience as
the longest walk observed is much smaller than the size of the network and shows no
change as that size is varied. Exponential decay 
of the random walk probabilities is not apparent for this network. 
This is in contrast to ER networks above the percolation threshold. 
In figure \ref{erdecay}, the curves for graphs with $\langle k\rangle$ below the 
threshold, do not display exponential decay, but this should be attributed to
a finite size effect from individual clusters. The BA network is simply connected
and investigations of the dependence on network size allow us to rule out
finite size as a cause of the curvature. 
Longer paths are more likely than would be expected from a fixed
exponential decay. This is due to the correlations and non homogeneous nature of the BA network. 
Krapivsky and Redner\cite{KrapivskyRedner} compute the age dependence of the 
degree distribution and support the notion of a highly connected old core to the network with
high degree nodes.
Long paths access this old core  and the 
random walks that access this region
tend to stay there and are less likely to be absorbed.

\subsection{Labelling according to ($k,l$)}

The framework afforded by labelling nodes by ($k,l$) is helpful 
to make predictions for BA networks despite reservations about its 
accuracy in the context of ER networks.
In this section we compute the probability 
that a randomly selected node on this BA network 
has degree $k$ and $l$ links to leaf nodes. 
In fact, to simplify notation, we compute $q_{kp}$ where $p = k-l$ is the number of links to non-leaf nodes.
By considering the growth process, we find a recursion relation for the expected number of labeled
nodes which when expressed as probabilities becomes.
\begin{equation}
\left(2k -p+2\right) q_{kp} = \left(k-1\right) q_{k-1p} + \left(k-p+1\right) q_{kp-1}
\end{equation}
When combined with initial conditions $q_{11} = n_1$ and $q_{kk+1}=q_{k0}=0$ this difference equations allows all higher terms to  be computed. However, an explicit solution requires some work.

Based on explorations for small values of $k,p$ we partially disentangle the two indices by writing.
\begin{equation}
q_{kp} = P_{p-1}(k) {(k-1)!\over (2k+1)!!}
-Q_{p-2}(k) {(k-1)!\over (2k)!!}
\end{equation}
Where $P_n(k)$ and $Q_n(k)$ are polynomials in $k$ of degree $n$.
These obey the following difference equations.
\begin{eqnarray}
\left(2k -p+2\right) P_{p-1}(k)&=& \left(2k+1\right) P_{p-1}(k-1) + \left(k-p+1\right) P_{p-2}(k)\\
\left(2k -p+2\right) Q_{p-2}(k) &=& \left(2k\right) Q_{p-2}(k-1) + \left(k-p+1\right) Q_{p-3}(k)
\end{eqnarray}
The first equation can be solved immediately to give.
\begin{equation}
P_{p-1}(k) = {2\over (p-1)!}{(k+p-2)!\over k!}\left(k+2(p-1)^2\right)
\end{equation}
While the second equation has the more complicated solution.
\begin{equation}
Q_{p-2}(k) = {4\over (k+1)}\sum_{j=1}^{j<1+k/2}
{(-1)^j\over (2j-1)(2j-3)}{1\over (j-1)!}
{(k+p-2j+1)!\over (p-2j)!(k-j+1)!}
\end{equation}
The resulting  $q_{kp}$ match numerical values obtained from generating many BA networks.
They also obey the following.
\begin{eqnarray}
\sum_{p=1}^k q_{kp} &=& n_k = {4\over k(k+1)(k+2)}\\
\sum_{p=1}^k (k-p)q_{kp} &=& e_{1k} = {2(k-1)(k+6)\over k(k+1)(k+2)(k+3)}\\
\sum_{p=1}^k (k-p)^2 q_{kp} &=&  {(k-1)(k^3+13k^2+42k -48)\over k(k+1)(k+2)(k+3)(k+4)}
\label{BAsums}
\end{eqnarray}

On BA networks there is only one connected component and there are no walks of length one step.
The probability of a randomly selected leaf node connecting to a $(k,p)$ node is.
\begin{equation}
{(k-p)\over n_1} q_{kp} 
\end{equation}
With this information we can now predict the probability of a walk of length 2 steps.
\begin{eqnarray}
p_2 &=&\sum_{k=2}\sum_{p=1}^k {(k-p)\over n_1} q_{kp}{(k-p)\over k} \nonumber\\
&=&{3\over 2}\sum_{k=2}{1\over k}\sum_{p=1}^k {(k-p)^2} q_{kp}\\
&=&{3\over 2}{1\over 144}(48\pi^2-419)= 0.570219\dots \nonumber
\label{BAlen2}
\end{eqnarray}
The first line exposes the separate factors for the transition and for the random walk,
while the second line shows that this can be expressed in
terms of the sums computed above in equation (\ref{BAsums}).
Numerical simulations on 100 graphs of size $10^5$ observe 
$0.57 \pm 0.02$, fitting the prediction well.

We can proceed to estimate the probability of walks of length 3 and over by 
extending this argument. However, this requires knowledge of the correlations between
the $(k,p)$ values of linked nodes. We have not computed this four index
quantity, but have made some estimates using the degree correlations in 
equation (\ref{BAcorrelate}).
These estimates are not accurate and we do not present them here.

Even though figure \ref{bawalk} clearly shows a curve,  
we can use (\ref{roughexp}) to estimate the survival probability.
We expect the analogy of $w_{kl}$, namely $p q_{kp}/\langle p\rangle$, 
to represent the transition probability to a ($k,l$) site. Here the mean number of connections to  non-leaf
nodes is given by $\langle p\rangle = 4/3$, so
the probability of making a further step without absorption is.
\begin{equation}
\sum_{k=2}\sum_{p=1}^k {p q_{kp}\over \langle p\rangle}
= 1- {q_{11}\over \langle p\rangle}
={1\over 2}
\end{equation}
The decay exponent that this argument suggests is therefore 
a constant $\gamma = \log 2$. We do not
show this prediction in figure \ref{bawalk} as it is not very accurate and we can
obtain a better prediction following the reasoning used for ER networks
and leading to the estimate of 
the survival probability given in equation (\ref{ERklbetterexp}).  
To do this we assume that  $p q_{kp}/\langle p\rangle$ can be regarded as an
occupation probability for non-leaf nodes. A factor of 2 is needed
for correct normalisation and probability of survival on the next step is.
\begin{equation}
\beta = \sum_{k=2}\sum_{p=1}^k {2 p^2 q_{kp}\over k\langle p\rangle}
={1\over 96}(48\pi^2-419)= 0.570219\dots
\end{equation}
We have not discovered any reason for the 
appearance of exactly the same factor as in equation (\ref{BAlen2}), and it seems to be a coincidence.


If we assume that the argument holds for all length paths, and
taking into account the lack of transience to normalise,
we can predict the following form.
\begin{equation}
p_t = {(1-\beta)\over \beta^2} \beta^t
\end{equation}
This form is shown in figure \ref{bawalk} and is provides a  
surprisingly good fit for shorter paths. As explained above, 
longer paths are affected by the correlations, enter the
region of old nodes, and are more likely than this approximation
predicts.

\subsection{Other approaches}

We have not attempted to extend the memoryless approximation along the lines of the
one described in section \ref{ERKN} for ER networks as the level of complication appears to be unwarranted.
Such an approximation would have to take into account the correlations between node
degrees and to do this needs to delve into the underlying directed nature of the BA graph.
The return probabilities would depend on the degree of the originating node.

\section{Conclusion}


This study of the
problem of random walks between leaf nodes of random networks
was initiated by the desire to model internet traffic, but it turns out that these walks 
probe network structure in a way that is not possible for
traditional non-absorbing walks. The mean first passage time
is a natural observable for non-absorbing walks, but to characterise
classes of graph it must be averaged over all pairs of nodes
and this reduces the information it can provide about the 
non-homogeneity of a network. 
By contrast, the random walks between leaves studied here provide a natural
subset of nodes over which the mean first passage time can be
averaged and this can give information about 
the heterogeneity of the network and
the extent to which edge nodes are connected to the rest of the network.
The clearest example of the way that walks between leaves can probe
network structure was the distinction between exponential decay of the
probability distribution of the length of the walk on homogeneous ER networks
and the non-exponential decay for heterogeneous BA networks. 
This however leads to a conundrum: the internet is hierarchical, yet the
probability distribution of time-to-live (TTL) values captured on the internet does not display any
systematic deviation from exponential.
This may indicate limits to the utility of random walks representing internet traffic.

Our analytic attempts to predict the probability of given length walks have relied on a
variety of techniques. For short paths we have been able to enumerate,
or devise approximations that allow such enumerations. 
For longer paths we have had some success with ER networks by
assuming that the equilibrium occupation probabilities of a node are 
proportional to the number of its non-leaf links.
Together with results for the proportion of $(k,l)$ nodes,
this approach gives some useful predictions for BA networks
but does not capture the non-trivial decay of the walk due to intrinsic
correlations in the network.


\vskip 0,5 truecm

\noindent{\bf Acknowledgment:} 
I am indebted to 
Dr B.~Ghita for collecting the TTL data for the internet that was the original motivation for this work.

 
\pagestyle{plain}


\baselineskip =18pt
\begin{thebibliography}{0}

\bibitem{internetdiffusion}
{B. Tadic and G. Rodgers, 
{\it Packet Transport on Scale Free Networks},
Advances in Complex Systems {\bf 5}, 445 (2002).}

\bibitem{internetdiffusion2}
{B. Tadic and S. Thurner, 
{\it Information Super-Diffusion on Structured Networks},
Physica {\bf A 332}, 566 (2004).}

\bibitem{weightedinternet}
{Y.~Zhang, S.~Zhou, Z.~Zhang, J.~Guan, S.~Zhou and G.~Chen,
{\it Traffic Fluctuations on Weighted Networks},
IEEE Circuits and Systems, {\bf 33}, (2012).}

\bibitem{ReigerNoh}
{J.D.~Noh and H.~Reiger,
{\it Random Walks on Complex Networks},
Phys. Rev. Lett. {\bf 92}, 118701 (2004).}


\bibitem{DenseER}
{V.~Sood, S.~Redner and D.~ben-Avraham,
{\it First Passage Properties of the Erd\H{o}s-R\'enyi Random Graphs},
J. Phys {\bf A38} 109-123 (2005).}

\bibitem{Olivier}
{O.C.~Martin and P.~Sulc,
{\it Return probabilities and hitting times of random walks on sparse
Erd\H{o}s-R\'enyi graphs}, Phys Rev {\bf E 81}, 031111 (2010).}

\bibitem{MFTER}
{A.~Baronchelli and V.~Loreto,
{\it Ring structures and mean first passage time in networks},
Phys. Rev. {\bf E 73}, 026103 (2006) .}

\bibitem{MasudaKonno}
{N.~Masuda and N.~Konno,
{\it Return times of random walk on generalized random graphs},
Phys Rev {\bf E 69} 066113 (2004).}

\bibitem{Redner}
{S.~Redner 
{\it A guide to first-passage processes}. Cambridge Univ. Press, New York, (2001).}


\bibitem{RandomGraph}
{B. Bollob\'as, 
{\it Random Graphs}, Academic Press, New York (1985).}
%

%

%

\bibitem{BarabasiAlbert}
{R. Albert and A. L. Barab\'asi,
{\it Statistical Mechanics of Complex Networks},
Rev. Mod. Phys., {\bf 74}, 47 (2002).}


\bibitem{Feller}
{W.~Feller {\it An Introduction to Probability Theory and its Applications},
Wiley, New York, (1950).}


\bibitem{GuptaSeth}
{H.C.~Gupta and Asha Seth,
{Random walk in the presence of absorbing barriers},
Proc National Institute of Sciences of India, {\bf 32}, 472-480,  (1966).}

\bibitem{ReedMolloy}
{M. Molloy and B. Reed, 
{\it A critical point for random graphs with a given degree sequence}, 
Random Structures and Algorithms {\bf 6}, 161-179 (1995).}



\bibitem{MasudaKonnoNote}
{Note that the combinatorial factor in Masuda and Konno's equation (17) is a Catalan number.}


%
\bibitem{GeneratingFunctions}
{M. E. J. Newman, S. H. Strogatz, D. J. Watts
{\it Random graphs with arbitrary degree distributions and their applications}
Phys. Rev. {\bf E 64}, 026118 (2001).}
%

\bibitem{KrapivskyRedner}
{P.L.~Krapivsky and S.~Redner, 
{\it Organization of growing random networks},
Phys. Rev. {\bf E 63}, 066123 (2001).}

 
 \bibitem{BAFiniteSize}
{B Waclaw, I.M Sokolov,
 {\it Finite size effects in Barab\'asi-Albert growing networks},
Phys. Rev. {\bf E 75}, 056114 (2007).}





%
%
\end{thebibliography}
\end{document}